\documentclass[aps,prd,amsmath,amssymb,superscriptaddress,nofootinbib,10pt,showpacs,twocolumn]{revtex4-2}

\usepackage[]{latexsym}
\usepackage[english]{babel}
\usepackage{graphicx}
\usepackage{hyperref}
\usepackage{color}
\usepackage{bm}

\usepackage{array} 

\allowdisplaybreaks

\newcommand{\ve}[1]{{\text{\bf #1}}} 
\newcommand{\vk}{\ve k}

\usepackage{booktabs}
\usepackage{float} 
\newcommand{\ra}[1]{\renewcommand{\arraystretch}{#1}}

\begin{document}

\title{Testing gravity with the full-shape galaxy power spectrum: First constraints on scale-dependent modified gravity}

\author{Alejandro Aviles}
\email{aviles@icf.unam.mx}

\affiliation{Instituto de Ciencias F\'isicas, Universidad Nacional Autónoma de México,  62210, Cuernavaca, Morelos, M\'exico}

\begin{abstract}
Since the discovery of the accelerated expansion of the Universe in 1998, modified gravity (MG) theories have attracted considerable attention as alternatives to dark energy (DE). While distinguishing the effects of MG from those of DE using the cosmic expansion alone is difficult, the large-scale structure is expected to differ significantly. Among the plethora of MG models, we are particularly interested in those that introduce a scale dependence in the growth of perturbations; specifically, theories that introduce fifth forces mediated by scalar fields with a finite range accessible to cosmological probes. This is the case with $f(R)$ gravity, which is widely regarded as the most studied model in cosmology. In this work, we utilize, for the first time, the full-shape power spectrum of galaxies to constrain scale-dependent modified gravity theories. By using BOSS DR12 dataset, along with a BBN prior on $ \omega_b $ and a Planck 2018 prior on $ n_s $, we obtain an upper bound of $ |f_{R0}| < 5.89 \times 10^{-6} $ at 68\% confidence level (c.l.) and $ < 1.53 \times 10^{-5} $ at 95\% c.l. for the Hu-Sawicki ($ n=1 $) model. We discuss that it is highly unlikely these constraints will be significantly improved by future galaxy spectroscopic catalogs, such as DESI and Euclid.
\end{abstract}

\maketitle


\begin{section}{Introduction}

The discovery of the accelerated expansion of the Universe marked a new era in cosmology \cite{SupernovaCosmologyProject:1998vns,SupernovaSearchTeam:1998fmf}. Although the simplest explanation for this phenomenon is provided by a cosmological constant, several groups have dedicated considerable effort to search for alternative descriptions that may ultimately expand our understanding of nature. One possibility is that general relativity (GR) is not the correct theory for describing gravity on large scales \cite{Deffayet:2001pu,Capozziello:2002rd,Dvali:2003rk,Carroll:2003wy,Sotiriou:2008rp}. Besides its impact over the background cosmology, if a modified gravity (MG) theory is responsible for the universe' accelerated expansion, one might expect to find its signatures in the clustering of matter as well \cite{Clifton:2011jh,Koyama:2015vza,Joyce:2016vqv,Ishak:2018his}. 

GR is distinctive in that, in the absence of anisotropic stresses, such as at late times, both scalar potentials of the metric in the Newtonian gauge are equal. In MG theories, however, this is generally not the case \cite{Bertschinger:2008zb}. Instead, the potential in the Poisson equation differs from the potential that sources the geodesic equation. Thus, while the universe's background evolution may not differentiate MG from dark energy (DE), large-scale structure (LSS) often differs. In DE models, LSS behaves similarly to $\Lambda$CDM as DE does not cluster. 
In contrast, typical MG theories introduce a finite-range attractive ``fifth-force''  mediated by a scalar field with mass $m_\phi$. The force between particles separated by distances larger than $1/m_\phi$ is exponentially suppressed, meaning that on the largest cosmological scales, gravity behaves according to GR. Successful MG models, however, employ screening mechanisms \cite{Vainshtein:1972sx,Khoury:2003rn,Hinterbichler:2010es,Brax:2010gi}, which arise from nonlinearities in the scalar field and drive the equations of motion back to those of GR. It is at intermediate scales, typically from several tens of megaparsecs upward, where cosmology can effectively search for modifications to gravity.

In the galaxy power spectrum, the signature will appear above comoving wave numbers $k \gtrsim k_\text{MG}(a) \equiv a m_\phi(a)$, where $a$ is the scale factor. Notice that the mass is dynamical, as it generally depends on environmental factors such as the local density or the local gravitational strength (e.g. \cite{Khoury:2003rn}). In a cosmological context, this is reflected in its dependence on the scale factor $a$. For some models, the mass either vanishes or is too small, such that the scale of onset of MG is even larger than the horizon. Such theories produce linear growth of density fluctuations that can be considered scale independent  for practical purposes. On the other hand, we refer to theories as scale dependent when $k_\text{MG}$ lies within the range accessible to cosmological probes.
Since the ``new'' force is attractive and scale dependent, the power spectrum is enhanced  above $k_\text{MG}$ in comparison to that in GR. This is the signature one would expect to observe, and use to constrain possible deviations from GR.

In this work we deal with the specific case of $f(R)$ gravity \cite{Buchdahl:1970ynr,STAROBINSKY198099}, that modifies the Einstein-Hilbert action by adding an arbitrary function of the Ricci scalar to its Lagrangian density, $R \rightarrow R + f(R)$. At the linear level in cosmological perturbation theory (PT), the gravitational strength modifies the evolution of density fluctuation, such that the linear growth function equation takes the following form
\begin{align} \label{Dplus}
    &\ddot{D}_+(k,t) + 2 H \dot{D}_+ 
    \nonumber\\ & \quad
    - \frac{3}{2}\Omega_m(a) H^2 \left( 1 + \frac{2 \beta^2 k^2}{k^2 +  k^2_\text{MG}(a)}\right) D_+ = 0,
\end{align}
with $2\beta^2=1/3$. We are interested in theories where $a m_\phi(a)$ grows as we back in time, since these reduce to GR in the early universe, and only affects the late times when the acceleration of the Universe takes places. We notice from Eq.~\eqref{Dplus} that the linear growth carries a $k$ dependence as a consequence of the (finite) mass of the scalar field that was introduced into the theory.  

The most studied $f(R)$ theory is the Hu-Sawicky (HS) gravity \cite{Hu:2007nk}. 
 The mass of the associated scalar field is given by 
 \begin{align}
     m_\phi(a) 
     &= \sqrt{\frac{3  H_0^2}{2 |f_{R0}|}} \frac{(\Omega_m a^{-3} + 4(1-\Omega_m))^{(2+n)/2}}{(4-3\Omega_m)^{n/2}},     
 \end{align}
 with $f_{R0}$ the derivative of $f(R)$ with respect to $R$ evaluated today, $H_0$ the Hubble constant and $\Omega_m$ the matter abundance at present time.

In Fig.~\ref{fig:K_MG_models} we show the onset scale of MG, given by $k_\text{MG}$, as a function of the redshift for the HS model. We choose different parameters $n=0.1,1.0,2.0$ with $f_{R0} = -10^{-5}$ fixed (top panel), and $f_{R0}=-10^{-4}, -10^{-5}, -10^{-6}$, with $n=1$ fixed (bottom panel). In all cases the mass $m_\phi(a)$ decays with time. Although a minima in $k_\text{MG}(a)$  appear because of the scale factor $a$ that converts physical to comoving wave numbers. 
The vertical lines serve to mark the redshift bins $z_1=0.38$ and $z_3=0.61$ of the Baryon Oscillation Spectroscopic Survey (BOSS) of the SDSS-III \cite{BOSS:2012dmf,BOSS:2016wmc}. Notice that in regions below $k_\text{MG}$, the LSS is insensitive to the MG effects. Hence, this plot suggests that the redshifts covered by BOSS are idoneous to test HS model with parameter $n\sim 1$. 
In the bottom panel we further show the horizontal (dashed) line that correspond to $k=0.2 \,h \,\text{Mpc}^{-1}$, which marks the onset of the breakdown  of PT. %
In this work we will fit BOSS data using the HS model, in part for the above mentioned reason, but also because there are several N-body state-of-the-art simulations (in particular the \texttt{MG-GLAM} simulations \cite{Hernandez-Aguayo:2021kuh,Ruan:2021wup}), that we already have used to test our PT methodology in Ref.~\cite{Rodriguez-Meza:2023rga}.

 \begin{figure}
 	\begin{center}
 	\includegraphics[width=2.5 in]{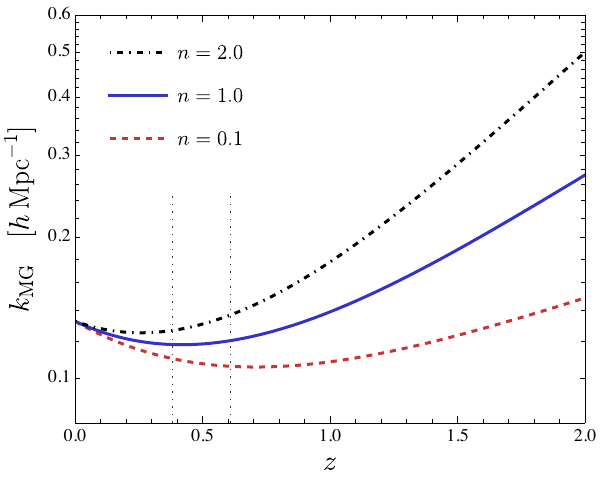}
 	\includegraphics[width=2.5 in]{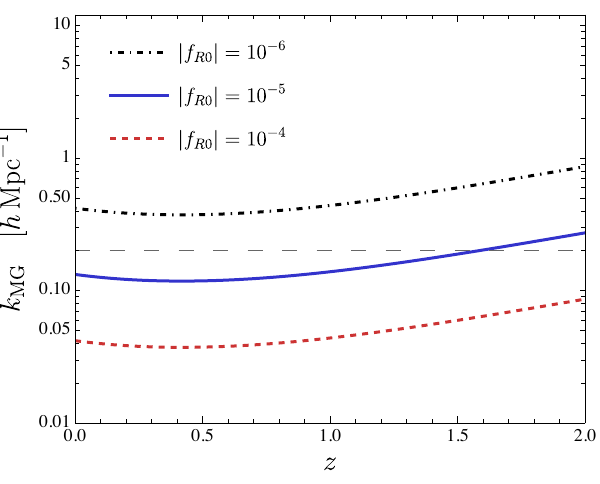}
 	\caption{Onset scale of MG,  $k_\text{MG}=a m_\phi(a)$, as a function of redshift $z$ for the $f(R)$ HS model. Top panel: we show curves for parameters $n=0.1,1.0,2.0$ with $f_{R0} = -10^{-5}$ fixed. The vertical lines serve to mark the redshift bins $z_1=0.38$ and $z_3=0.61$ of BOSS. \textit{Bottom panel}: $f_{R0}=-10^4, -10^5, -10^6$, with $n=1$. We depict the scale $k=0.2 \,h \,\text{Mpc}^{-1}$ (horizontal-dashed line), marking the failure of PT.
 	\label{fig:K_MG_models}}
 	\end{center}
 \end{figure}


\end{section}

\begin{section}{Theoretical model}\label{sect:theory}

In this section, we provide a brief overview of the theory used for fitting the BOSS galaxy power spectrum. For details, see Ref.~\cite{Rodriguez-Meza:2023rga}.
We utilize the PT for LSS formation, which has been developed for over more than two decades  \cite{Bernardeau:2001qr,McDonald:2006mx,McDonald:2009dh,Matsubara:2007wj,Baumann:2010tm,Carlson:2012bu,Vlah:2015sea,Chen:2020fxs}, and by now has been largely applied to BOSS data sets; see Refs.~\cite{Ivanov:2019pdj,DAmico:2019fhj,Chen:2021wdi,Chen:2024vuf,Noriega:2024lzo}, among many others. However, most of this work has been devoted to scenarios close to $\Lambda$CDM, employing the so-called Einstein-de Sitter (EdS) approximation that assumes $f(t)= \Omega_m^{1/2}(t)$, where $f(t)$ is the scale independent growth rate obtained from Eq.~\eqref{Dplus} without the effects of MG. 
In this work, we go beyond EdS and utilize the PT developed in \cite{Aviles:2017aor,Aviles:2020wme}, based on works by other several authors \cite{Koyama:2009me,Taruya:2013quf,Brax:2013fna,Bellini:2015oua,Taruya:2014faa,Taruya:2016jdt,Winther:2017jof,Fasiello:2017bot,Bose:2016qun,Bose:2017dtl,Bose:2018orj,Valogiannis:2019xed,Valogiannis:2019nfz,Aviles:2018qotF,Aviles:2018saf}. While this theory is comprehensive, its practical application is limited, as the numerical implementation is highly time consuming, making parameter space exploration infeasible. Therefore, we employ the \texttt{fk}-PT  approximation \cite{Aviles:2021que,Noriega:2022nhf,Rodriguez-Meza:2023rga}, that keeps the scale dependence in the growth factors, arising from the fact that, at linear order in PT, the velocity divergence and the density fluctuation of matter fields are not equal, but\footnote{We follow the definition $\theta = \nabla \cdot \ve v / (a H f_0)$ with $\ve v$ the peculiar velocity field.}  
\begin{equation}
    \theta^{(1)}(\vk,t) = \frac{f(k,t)}{f_0(t)}\delta^{(1)}(\vk,t),
\end{equation}
where the scale-dependent growth rate is
\begin{equation}
    f(\vk,t) \equiv \frac{d \log D_+(k,t)}{d \log a(t)}
\end{equation}
and $f_0(t) = f(\vk=0,t)$ is the growth rate at the largest scales.\footnote{When the range of the fifth force becomes very large, the theory becomes scale-independent. This is the case of the normal branch of  Dvali-Gabadadze-Porrati (DGP) gravity \cite{Dvali:2000hr}, at linear order. This theory was constrained using BOSS data in Ref.~\cite{Piga:2022mge}. Scale-independent  PT/EFTs for MG has been studied in several other works, e.g., \cite{Cusin:2017wjg,Bose:2018orj}.} Because the growth rate is $\vk$ dependent, the linear power spectra of the matter velocity and density fields are not equal, but they become related by
\begin{align}
    P_{\delta\theta}^L(k) &= \frac{f(k)}{f_0} P_{\delta\delta}^L(k), \quad   
    P_{\theta\theta}^L(k) = \left( \frac{f(k)}{f_0} \right)^2 P_{\delta\delta}^L(k).
\end{align}

The one-loop redshift ($s$) space galaxy power spectrum is given by
\begin{equation}\label{pofk}
   P_s(k, \mu) = P_s^\text{PT}(k, \mu) + P_s^\text{EFT}(k, \mu) + P_s^\text{shot}(k, \mu),
\end{equation}
where $\mu$ is the cosine of the angle between the line of sight direction and the wave vector $\vk$.

The ``pure'' perturbative term is given by
\begin{align}\label{PsME}
   P_s^\text{PT}(k, \mu) &= P_{\delta\delta}(k) + 2 f_{0} \mu^2 P_{\delta\theta}(k) + f_{0}^2 \mu^4 P_{\theta\theta}(k) \nonumber\\
   &+ A^\text{TNS}(k,\mu) + D(k,\mu),  
\end{align}
where the one-loop real space power spectra $P_{\delta \delta}$, $P_{\delta\theta}$, and $P_{\theta\theta}$ are displayed in Eqs.~(3.74)-(3.76) of \cite{Rodriguez-Meza:2023rga}. The functions $A^\text{TNS}$, defined in \cite{Taruya:2010mx}, and $D(k,\mu)$ are given by Eqs.~(3.51) and (3.53) of \cite{Rodriguez-Meza:2023rga}. 

We further add effective field theory (EFT) counterterms of the form
\begin{equation}
    P_s^\text{EFT}(k,\mu) = \big(\alpha_0 + \alpha_2 \mu^2 + \cdots \big) k^2 P_L(k),   
\end{equation}
accounting the backreaction from the unmodeled small scales---out of the reach of PT---over the large scales. The shot noise term is
\begin{equation}
    P_s^\text{shot}(k, \mu) = 
    \alpha_0^\text{shot} + (k \mu)^2 \alpha^\text{shot}_2. 
\end{equation}

The biasing scheme differs slightly from that of $\Lambda$CDM, and is presented in \cite{Aviles:2020wme}. Indeed, for scale-dependent MG, this scheme is not complete at any order in PT, but higher derivatives operators enter into the theory (see also \cite{Desjacques:2016bnm}). However, we exclude these terms because they become degenerate with the EFT counterterms.

The final step is to model the spread of the BAO oscillations caused by large-scale bulk flows. To achieve this, we employ IR resummations, as outlined in Refs.~\cite{Senatore:2014via, Vlah:2015sea}. The details of our implementation, closely following Refs.~\cite{Ivanov:2018gjr, Ivanov:2019pdj}, are presented in Sec.~3.6 of \cite{Rodriguez-Meza:2023rga}. This approach allows us to compute the final expression for the power spectrum, \( P_s^\text{IR}(k,\mu) \). For fitting the data, we use the multipoles from the equation
\begin{equation}\label{Pells}
P_\ell(k) = \frac{2 \ell + 1}{2} \int_{-1}^{1} d\mu \, P_s^\text{IR}(k,\mu) \mathcal{L}_{\ell}(\mu),    
\end{equation}
with $\mathcal{L}_{\ell}$ the Legendre polynomials.

The one-loop power spectrum is computed in a fraction of a second using the code \texttt{fkpt}.\footnote{\href{https://github.com/alejandroaviles/fkpt}{https://github.com/alejandroaviles/fkpt}} The code receives as input the $\Lambda$CDM linear power spectrum and rescales it with the use of the growth function to obtain the MG linear power spectrum
\begin{equation} \label{PLrescaling}
    P_L^\text{GR}(k,z) =   \left( \frac{D_+^\text{GR}(k,z)}{D_+^\text{$\Lambda$CDM}(k,z_0)} \right)^2
    P_L^\text{$\Lambda$CDM}(k,z_0).
\end{equation}
This equation holds for models that modify gravity only at high redshifts. In the case of HS, we have demonstrated that it is valid to an excellent approximation when compared to results obtained using \texttt{hi\_class} \cite{Zumalacarregui:2016pph}. For models where this is not true, one can use directly the MG power spectrum computed with a different code.

\end{section}

\begin{section}{Data}\label{sect:setting}

The BOSS-DR12 dataset spans redshifts from 0.2 to 0.75 over an effective area of 9,329 $\text{deg}^2$ and volume of $18.7 \, \text{Gpc}^3$. These are organized into bins with effective redshifts at $z_1 = 0.38$, $z_2 = 0.51$, and $z_3 = 0.61$. For our study, we utilize the two nonoverlapping bins, $z_1$ and $z_3$. Additionally, each redshift bin is divided into two subgroups based on the galactic hemisphere in which the galaxies were observed, labeled as the ``North Galactic Cap'' and ``South Galactic Cap''. 

We use the products provided in \cite{Beutler:2021eqq}.\footnote{\href{https://fbeutler.github.io/hub/deconv_paper.html}{https://fbeutler.github.io/hub/deconv\_paper.html}} These include data vectors and window functions for BOSS-DR12 \cite{BOSS:2015zan,BOSS:2015ewx}, as well as the covariance matrices obtained from the MultiDark-Patchy mocks \cite{Kitaura:2015uqa}. We refer the reader to Ref.~\cite{Beutler:2021eqq} for details.

\begin{center}
\begin{table}
\ra{1.3}
\begin{center}
\begin{tabular} { l  c }

 Parameter &  Prior  \\
\hline
\vspace{0.15cm}

$\quad |f_{R0}|$ &  $\mathcal{U}(0,0.1)$ \\ 
\vspace{0.15cm}

$\quad \omega_c$ & $\mathcal{U}(0.05, 0.2)$\\
\vspace{0.15cm}

$\quad h$ & $\mathcal{U}(0.4,0.9)$   \\
\vspace{0.15cm}

$\quad \ln(10^{10}A_s)$ & $\mathcal{U}(2.0,4.0)$ \\
\vspace{0.15cm}

$\quad \omega_\text{b}$ & $\quad \mathcal{N}(0.02230,0.00038^2) \, $\\
\vspace{0.15cm}

$\quad n_s$ & $\quad \mathcal{N}(0.9649,0.0042^2) \, $\\
\vspace{0.15cm}

$\quad b_1$ & $\mathcal{U}(0.01, 10)$ \\
\vspace{0.15cm}

$\quad b_2$ & $\mathcal{U}(-10,10)$ \\
\vspace{0.15cm}

$\quad b_{s^2}$ & $\quad \mathcal{N}(-4/7(b_1 - 1),0.05^2) \, $\\
\vspace{0.15cm}

$\quad \alpha_0$ & $\mathcal{N}(0,20^2)$ \\
\vspace{0.15cm}

$\quad \alpha_2$ & $\mathcal{N}(0,20^2)$ \\
\vspace{0.15cm}

$\quad \alpha_0^\text{shot}$ & Uninformative \\
\vspace{0.15cm}

$\quad \alpha_2^\text{shot}$ & Uninformative \\

\hline

\end{tabular}
\caption{Parameters and their priors. $\mathcal{N}(\mu,\sigma^2)$ denotes a Gaussian prior with median $\mu$ and variance $\sigma^2$, while $\mathcal{U}(a,b)$ is uniform prior over the interval $(a,b)$. 
}
\label{table:priors}
\end{center}
\end{table}
\end{center}

\end{section}

\begin{section}{Results}\label{sect:analysis}

 \begin{figure}
 	\begin{center}
 	\includegraphics[width=3.0 in]{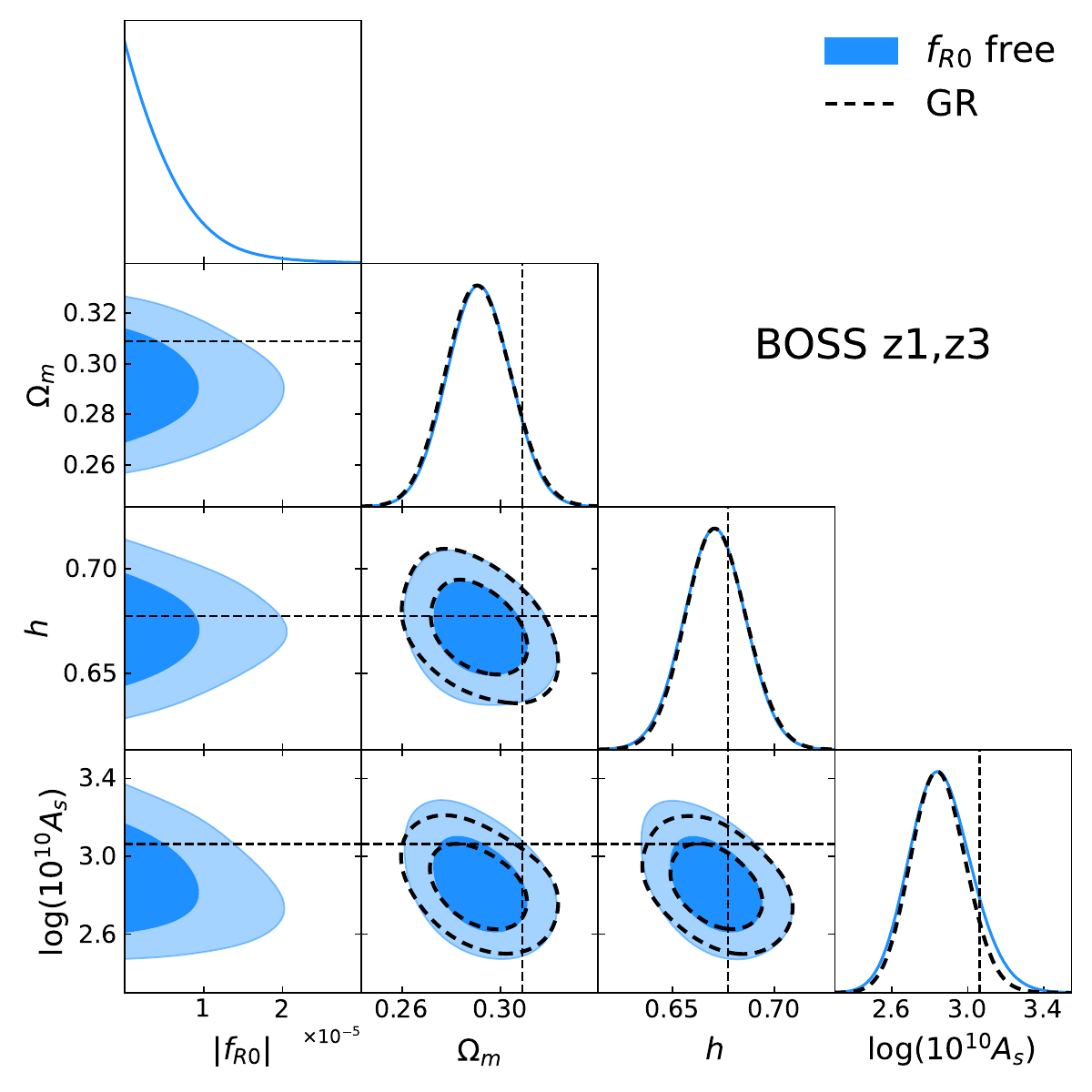}
 	\caption{Corner plots for BOSS $z_1$ and $z_3$ constraints on the HS $f(R)$ gravity (blue) and GR (black dashed). We show the 68\% and 95\% contours, as well as the marginalized one-dimensional distributions. The horizontal and vertical lines denote the best-fit values found by Planck 2018 \cite{Planck:2018vyg}.
 	\label{fig:constraints}}
 	\end{center}
 \end{figure}

In this section we confront the $f(R)$ HS ($n=1$) model with the BOSS data. To do this we run Monte Carlo Markov chains using the codes \texttt{CLASS} \cite{Blas:2011rf}, 
\texttt{fkpt}, and the affine-invariant sampler \cite{2010CAMCS...5...65G} implemented in \texttt{emcee} \cite{ForemanMackey:2012ig}. 
To analyze the chains and plot our results we utilize \texttt{GetDist}  \cite{Lewis:2019xzd}. 

The free parameters in our analysis are the primordial amplitude $\log(10^{10} A_s)$, the cold-dark matter density $\omega_c$, the reduced Hubble constant $h$ and the MG parameter $f_{R0}$. Additionally, we impose a Gaussian prior on the spectral index $n_s$ based on Planck best-fit \cite{Planck:2018vyg}, as well as a big bang nucleosynthesis (BBN) Gaussian prior for the baryon density $\omega_b$ \cite{Aver:2015iza,Cooke:2017cwo}. The neutrino mass is kept fixed at $\sum m_{\nu} = 0.06\, \text{eV}$. We utilize a set of nuisance parameters for each redshift bin and each galactic cap, encompassing the biases $b_1$, $b_2$ and $b_s$, the EFT counterterms $\alpha_0$, $\alpha_2$, and the noise parameters $\alpha_0^\text{shot}$ and  $\alpha_2^\text{shot}$. 

We analytically marginalize over EFT and noise parameters, which are linear in the power spectrum (before IR resummations); see, e.g. Sec.~3.4 of \cite{DAmico:2019fhj}. Hence, we never fit them. That is, in total we vary 18 parameters. 
A summary of these parameters and their priors is provided in Table \ref{table:priors}. 

The pipeline outlined here was validated in \cite{Rodriguez-Meza:2023rga} with the use of the $\Lambda$CDM \texttt{NSeries} mocks \cite{BOSS:2016wmc} and the $f(R)$ \texttt{MG-GLAM} simulation. In that study, we employed coevolution theory for biases $b_s$ and $b_{3nl}$. However, we give a small liberty to the tidal bias around its coevolution value, while the third-order nonlocal bias remains fixed to $b_{3nl} = 32/315(b_1 - 1)$ \cite{Chan:2012jj,Baldauf:2012hs,Saito:2014qha}. 

Since we are neglecting the nonlinearities responsible for producing the screening effect---that have been shown to be accurately handled by the EFT counterterms in \cite{Rodriguez-Meza:2023rga}---our analysis is limited to a moderate maximum wave number of $k_\text{max} = 0.17 \, h \,\text{Mpc}^{-1}$. Further, in \cite{Rodriguez-Meza:2023rga}, we found this choice to be optimal for our study. The minimum wave number is $k_\text{min} = 0.02 \, h \,\text{Mpc}^{-1}$.

Our preliminary results indicate that leaving the EFT counterterms uninformative makes it infeasible to place meaningful constraints on $f_{R0}$, likely due to the lack of sufficient information to break parameter degeneracies. 
 However, using very broad uniform priors (without analytical marginalization) does yield viable results. Therefore, we adopt Gaussian priors with standard deviation $\sigma = 20 \, h^{-1} \text{Mpc}$ for the EFT parameters. 

We present our results for the HS and GR models in Fig. \ref{fig:constraints} and Table \ref{Table:bestfits}. Notably, there are no significant degeneracies between $f_{R0}$  and the other cosmological parameters.  This is very likely due to the scale dependence of the gravitational strength. In scenarios where the large scales are impacted, as seen in the nDGP model in \cite{Piga:2022mge}, degeneracies tend to be very pronounced. The absence of such degeneracies in our case suggests that we do not encounter additional projection (or prior volume) effects than those present in the $\Lambda$CDM model \cite{Simon:2022csv,Hadzhiyska:2023wae,Noriega:2024lzo}, 
which may arise when too many parameters control the overall amplitude of the power spectrum ($\Omega_m$, $h$, $A_s$, $b_1$, and $n_s$ to some extent). In contrast, $f_{R0}$ is a parameter that influences directly, and uniquely, the shape of the power spectrum at scales above $k_\text{MG}$. 

At 1 and 2$\sigma$ we obtain the following constraints for $|f_{R_0}|$ in the HS, $n=1$ model are
\begin{align}
    |f_{R0}| &< 5.89\times 10^{-6}   \qquad   (68 \% \,\,\text{c.l.}), \label{1sigma}\\
    |f_{R0}| &< 1.53\times  10^{-5}  \qquad   (95 \% \,\,\text{c.l.}). \label{2sigma}
\end{align}
 In the next section, we will discuss why it is highly unlikely that these constraints can be significantly improved using the power spectrum from galaxy surveys. We will also compare our result with others in the literature.

\begin{table}
\centering
\setlength{\tabcolsep}{0.80em} 
\renewcommand{\arraystretch}{1.4}
\begin{tabular}{llll}
\begin{tabular}{*3c}\toprule
  \textbf{Parameter}                                        & $f_{R0}$ free             & GR \\\midrule
 $h$                                                        & $0.671\pm 0.014$        & $0.671\pm 0.014$\\
 $\Omega_{m}$                                             & $0.291\pm 0.012$        &  $0.291\pm 0.013$\\
 $f_{R0}$                                                   & $<5.89\times 10^{-6}$     &   $--$\\
 $\log(10^{10}A_s)$                                         & $2.86^{+0.13}_{-0.17}$ &   $2.85\pm 0.14$\\
 $b_1$                                                      & $2.19^{+0.19}_{-0.16}$          &   $2.19^{+0.17}_{-0.14}$\\\bottomrule
\end{tabular}
\end{tabular}
\caption{\label{Table:bestfits} 1-dimensional BOSS $z_1$ and $z_3$ 0.68 c.l. intervals for the HS $f(R)$ gravity and GR.}
\end{table}

\end{section}

\begin{section}{Discussions and prospects for future surveys}\label{sect:discussion}

Nowadays, the most powerful tool to test the galaxy power spectrum from spectroscopic surveys is the full-shape technique based on PT. However, the scales accessible through this theory are limited to values below $k\sim 0.20 \,h\,\text{Mpc}^{-1}$. In Fig.\ref{fig:K_MG_models} we showed such scales to give us a sense of the limits of our method. In the bottom panel of that figure, the horizontal dashed lines show this PT maximum scale. It is noteworthy that this lies well below  $k_\text{MG}$ for $|f_{R0}|=10^{-6}$, indicating that the upper bounds for $f_{R0}$ imposed by BOSS are close to the limits of what can be derived using PT. The situation is more critical as our approach assumes that the fifth-force screening is absorbed by EFT counterterms, limiting our analysis to $k < 0.17 \,h\,\text{Mpc}^{-1}$. However, it is possible that different techniques to treat screenings, as the one presented in \cite{Euclid:2023bgs}, can reach higher wave numbers and obtain slightly tighter constraints. 

 \begin{figure}
 	\begin{center}
 	\includegraphics[width=3.02 in]{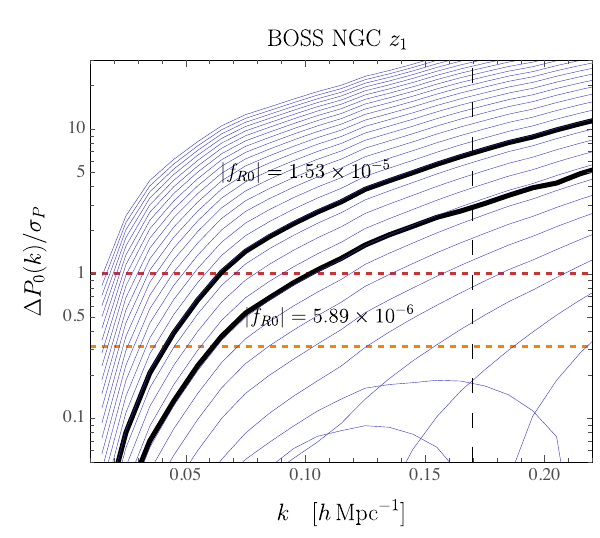}
 	\includegraphics[width=3.0 in]{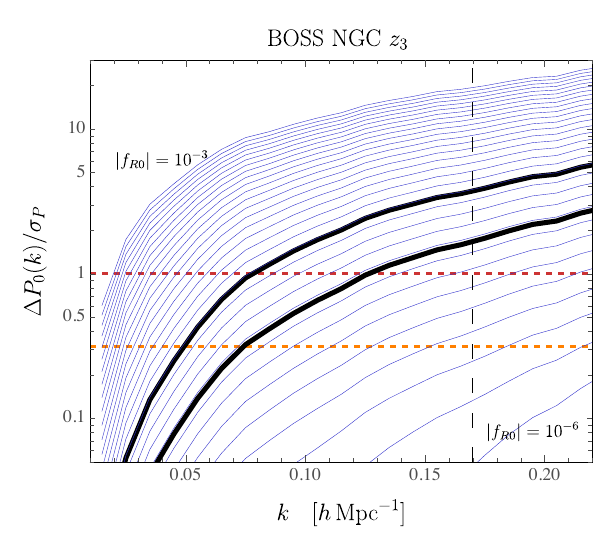}
 	\caption{Signal-to-noise ratio (S/N) given by the difference in the monopole of the power spectrum divided by the errors $\sigma_P$. The lower vertical line (orange-dashed) shows the S/N rescaled with a factor $\sqrt{10}$, roughly representing a volume 10 times that of BOSS.
 	\label{fig:StoN}}
 	\end{center}
 \end{figure}

Expanding on this idea, Fig.\ref{fig:StoN} presents plots illustrating the signal-to-noise ratio (S/N), as given by the difference in the monopole of the power spectrum, $\Delta P_{\ell=0}= P_0^\text{MG} - P_0^\text{$\Lambda$CDM}$, divided by the errors $\sigma_P$ obtained from the MultiDark-Patchy mocks. We show the $z_1$ (upper panel) and $z_3$ (bottom panel) North Galactic Cap samples. From top to bottom, the blue thin lines show models with MG parameters from $f_{R0}=-10^{-3}$ to $-10^{-6}$, spaced in logarithmic intervals. The solid thick black lines show our constraints to 1 and $2\sigma$ given in Eqs.~\eqref{1sigma} and \eqref{2sigma}. The red dashed horizontal lines indicate the S/N $=1$ threshold, which helps visualize which models can be distinguished, in the best-case scenario, using the BOSS-DR12 dataset. Due to mode coupling in nonlinear theory, there can be small MG signatures even below $k_\text{MG}$. 
However, these are too small, such that even with a considerable rescaling of the effective surveyed volumes, the errors remains sufficiently large and no significant improvements are expected: the $1\sigma$ constraints will lie well above $|f_{R0}| = 10^{-6}$. This is shown with the orange dashed horizontal lines in both panels, where the effective volume of the surveys are multiplied by a factor of 10.
This analysis demonstrates that the limits we derive for $|f_{R0}|$ are both expected and reasonable. Furthermore, it indicates that substantially tightening these constraints using the full-shape technique alone in galaxy counts is unlikely.

Finally, our constraints should be contrasted with those obtained using other methods. Currently, the most restrictive upper limit in the HS model is an impressive $|f_{R0}| < 10^{-8}$, obtained from astrophysical scales \cite{Landim:2024wzi}, exceeding by far any limit imposed by cosmological probes. The tightest cosmological constraint in the HS model comes from a (very) recent joint analysis of weak-lensing mass-calibrated South Pole Telescope clusters and \textit{Planck} 2018 data, yielding $|f_{R0}| < 10^{-5.32} = 4.79 \times 10^{-6}$ \cite{SPT:2024adw}. Another recent cluster abundance analysis, conducted by eROSITA alone, provides a weaker constraint of $ |f_{R0}| < 10^{-4.12} = 7.59 \times 10^{-5} $ \cite{Artis:2024eco}. From LSS observations, the most stringent  limit  is given by weak lensing peak statistics from CFHTLenS \cite{Liu:2016xes}, which yields an upper bound of $ |f_{R0}| < 10^{-4.82} = 1.51 \times 10^{-5} $, closely matching our result. But by including CMB priors on $ \Omega_m $ and $ A_s $, the authors achieve significantly tighter constraints, falling below $ 10^{-5} $.

\end{section}

\begin{section}{Conclusions}\label{sect:conclusions}

In this work, we have confronted the HS $ f(R) $ model using LSS data from the BOSS-DR12 of the SDSS-III survey. We employed a full-shape analysis to constrain the $ |f_{R0}| $ parameter. One of the challenges in such analysis is that using the standard EdS kernels in PT is unreliable. To address this, we utilized the fk-PT method developed in recent papers. Although this approach is not fully complete and employs a simplified version of the \textit{full} kernels, it has proven to be an excellent approximation in $ f(R) $ gravity. The pipeline we used here was recently validated through simulations, with mocks designed for both MG and the $\Lambda$CDM model.

By adding a BBN prior on $ \omega_b $ and a Planck 2018 prior on $ n_s $, we placed an upper bound of $ |f_{R0}| < 1.53 \times 10^{-5} $ at 95\% confidence level. This result represents the first constraint on scale-dependent MG using the full-shape galaxy power spectrum.

An interesting aspect of our results is that introducing the parameter $ |f_{R0}| $ does not seem to impact the rest of cosmological parameters. We believe this occurs because the MG signature does not substantially change the large-scale power spectrum or its overall amplitude. In contrast, what we refer to in this work as scale-independent MG theories, such as DGP, introduce new parameters that strongly influence the overall power spectrum as much as other cosmological parameters, such as $ A_s $, $ h $, $ \Omega_m $, as well as the linear bias $ b_1 $. On the other hand, $|f_{R0}|$ influences the shape of the power spectrum above a characteristic wave number $ k_\text{MG} $, leaving the larger scales unaffected. This property does not seem to introduce additional projection effects beyond those already present in the $ \Lambda$CDM model, leading us to believe that scale-dependent MG models are more easily constrained than scale-independent ones, despite the more intricate PT required.

We also discussed the expected constraints from future full-shape analyses of spectroscopic galaxy surveys, concluding that substantial improvements are unlikely due to the limitations of PT, as the results obtained with BOSS data are already at the edge of this limit. However, incorporating weak lensing data could provide additional constraints since the lensing potential depends directly on the sum of the two gravitational potentials in the metric (in Newtonian gauge). We plan to address such a joint analysis in future publications.

\end{section}

\acknowledgments

I would like to thank Arka Banerjee, Jorge L. Cervantes-Cota, Baojiu Li, Gustavo Niz, Hernan E. Noriega, Mario A. Rodriguez-Meza and Georgios Valogiannis.  
 
This work is partially supported by CONAHCyT grant CBF2023-2024-162 and PAPIIT IG102123.

\appendix

 \bibliographystyle{JHEP} 
 \bibliography{refs.bib}

\end{document}